\documentclass[10pt,letterpaper]{article}
\usepackage{opex3-mod}

\def\Ve#1{\mbox{\boldmath $#1$}}

\begin{document}

\pagestyle{plain}

\title{Deep Learning of Diffuse Optical Tomography based on Time-Domain Radiative Transfer Equation}
\author{Yu-ichi Takamizu$^{1}$,
Masayuki Umemura$^{1}$, 
Hidenobu Yajima$^{1}$, 
Makito Abe$^{1}$, and
Yoko Hoshi $^{2}$}

\address{$^{1}$ 
Center for Computational Sciences, University of Tsukuba, 
1-1-1 Tennoudai, Tsukuba-shi Ibaragi Japan \\
$^{2}$ 
Preeminent Medical Photonics Education and  Research Center, Hamamatsu University School 
of Medicine, 1-20-1 Handayama, Higashi-ku Hamamatsu, Sizuoka Japan }

\email{takamizu@ccs.tsukuba.ac.jp}



\begin{abstract}
Near infrared diffuse optical tomography (DOT)  provides an imaging modality for the oxygenation of tissue.  
In this paper, we propose a novel machine learning algorithm based on time-domain radiative transfer equation. We use temporal profiles of absorption measure for a two-dimensional model of target tissue, which are calculated by solving time-domain radiative transfer equation.
Applying a long-short-term memory (LSTM) deep learning method, we find that we can specify positions of cancer cells with high accuracy rates. We demonstrate that the present algorithm can also predict multiple 
or extended cancer cells. 
\end{abstract}
\section{Introduction}
Diffuse optical tomography (DOT) using near infrared light (700 nm- 900 nm) is one of the most sophisticated optical imaging techniques for biological tissue. 
Especially, DOT can provide a promising imaging modality for cancer detection owing to its contrast to hemoglobin oxygenation level. 
To reconstruct optical properties of tissue in DOT, 
two mathematical problems should be solved, i.e., 
the forward and the inverse problem  \cite{Hoshi16}. 
The  forward problem is to pursue the propagation of scattered light in biological tissue and thereby
predict the scattered light measurements for a setup of tissue optical properties, while
the inverse problem is to reconstruct tissue optical properties from scattered light measurements
by use of the forward model. The image reconstruction in DOT is a nonlinear ill-posed inverse problem,
which is subject to the shortage of data diversity and instabilities to noise. 
 Hence, the feasibility of DOT depends upon how precisely the  forward problem is calculated and
 how stably the inverse problem is solved. 
 
 The propagation of light with scattering and absorption is rigorously governed by 
 the radiative transfer equation (RTE) \cite{Klose10}. 
Numerical schemes to directly solve RTE in  biological tissue have been
proposed \cite{Klose99,Gonzalez09}. However, they are computationally expensive, since RTE in three-dimensional space results in a six-dimensional problem in a photon phase space.
Also, to converge the inverse problem, the forward problem should be solved several times for a single reconstruction. Hence, the RTE calculations have been an unavoidable bottleneck in algorithms 
to solve the inverse problem. 
So far, light propagation has been often approximated by a diffusion equation (DE)
using the P1 approximation of the RTE. 
The diffusion approximation is a simplification of the RTE for optically thick
media in which multiple scattering is dominant \cite{Wang07}.
Based on frequency-domain as well as time-domain DE, various approaches for image reconstruction in DOT have been attempted \cite{Pogue97,Okawa11}. 

However, actual tissue systems follow the so-called mesoscopic scattering regime, 
where light undergoes multiple scattering but the scattered light is not perfectly diffusive.
For mesoscopic scattering, the diffusion approximation breaks down 
especially near sources and boundaries. To circumvent the shortage of diffusion approximation,
hybrid schemes that combine radiative transfer with diffusion have been implemented \cite{Kim10,Tarvainen11,Fujii14}. 
Hybrid schemes can reduce dramatically the computational cost to solve light
propagation, but the boundaries between RTE and DE cannot be determined {\it a priori}.
Therefore, overall RTE calculations are desirable.  
As for frequency-domain RTE, several solvers of the inverse problem 
have been proposed \cite{Ren06,Gonzalez15}.
However, image reconstructions based on time-domain RTE are still immature. 
The time-domain RTE provides more functional information on biological tissue.

Recently, machine-learning has brought a new possibility for image reconstructions. 
Actually,  a deep learning scheme based on DE has been launched \cite{Yoo20}.  
In this paper, we construct a novel machine learning algorithm based on time-domain RTE. 
In this algorithm, training data are the temporal profiles of light propagation in 
models mimicking biological tissue. 
For the purpose, we have developed a time-domain RTE solver,
{\sf TRINITY} (Time-dependent Radiation Transfer in Near-infrared Tomography),
which is based on our former steady-state RTE solver, {\sf ART} \cite{Iliev06}.
As for machine learning, we utilize an LSTM (Long Short-Term Memory) method,
which is an extension of artificial recurrent neural network (RNN) architecture 
to process not only single data points but also entire sequences of temporal data.

\section{Methodology}

\subsection{Model}

 Throughout this paper, we set up a two-dimensional model. 
 We suppose target tissue with the size of 4cm$\times$4cm, in which
we allocate 28 domains to specify the positions of cancer cells, as
shown in Fig. 1. 
We assume a round absorber with diameter 5mm as a model of a cancer cell. 
This setup is based on the experiment using a phantom composed of
polyoxymethylene.
The optical properties of the phantom are characterized by 
\begin{equation}
\mu_a=0.21/{\rm cm}\,, \ \mu_s=22.45/{\rm cm}\,
\end{equation}
where $\mu_a$ and $\mu_s$ are the absorption and scattering coefficient, respectively.
Within this background material, an absorbing pole with following coefficients 
\begin{equation}
\mu_a=0.64/{\rm cm}\,, \ \mu_s=22.63/{\rm cm}\,,
\end{equation}
is inserted.
Scattering dominates absorption in the inserted pole as well as the background,
but the absorption coefficient is larger by a factor of three in the pole. 
Therefore, absorption features due to the pole emerge in outgoing light at the boundary.
That is a principle for detecting cancer positions in biological tissue.
The incident beams and detectors are located 
at eight points as shown in the Fig. 1, where the positions of incident beams are
labelled with S1-S8 and positions of detectors are with D1-D8.

We solve the following radiative transfer equation in two-dimensional space
to pursue scatted and absorbed light, 
\begin{equation}
\frac {1}{c}\frac{\partial I}{\partial t}+\Ve{n}\cdot \nabla I= -\mu I +\eta\,,
\label{RTE}
\end{equation}
with
\begin{equation}
\mu=\mu_a +\mu_s\,,\\
\eta=\mu_s \oint \phi(\Ve{n},\Ve{n}')I(\Ve{n}')d\Omega'\, ,
\end{equation}
where $I$ is the specific intensity of light, $\eta$ is the emissivity by scatted photons,
and $\phi$ is a phase function. 
The emissivity term represents 
radiative transfer from one direction $\Ve{n}'$ to another direction $\Ve{n}$. 
We employ the Henyey-Greenstein function for $\phi$, that is, 
\begin{equation}
\phi(\Ve{n},\Ve{n}')=
\frac{1}{2\pi}\frac{1-g^2}{1+g^2- 2g (\Ve{n}\cdot \Ve{n}')}\,,
\end{equation}
where $g$ is the scattering anisotropy parameter. 
The case of $g=1$ or $g=0$ means perfectly forward scattering and 
isotropic scattering, respectively. We take $g=0.62$ for the phantom. 

To solve Eq. (\ref{RTE}), we have developed a new solver,
{\sf TRINITY} (Time-dependent Radiation Transfer in Near-infrared Tomography)(Yajima et al. in prep),
which is based on our former steady-state RTE solver, {\sf ART} 
(Authentic Radiative Transfer) \cite{Iliev06}.
This method is based on an idea 
that light ray is designed independently of spatial grids.  
The significant advantage is simultaneous reduction of computational cost 
and numerical diffusion. 
We set up $64^2$ grids for radiative transfer calculations.

Using radiative transfer calculations, 
we define absorption measure as
\begin{equation}
A(t)={I_{\rm abs}(t) /I_{\rm noabs}(t)}-1\,,
\end{equation} 
where $I_{\rm noabs}$ is the intensity of outgoing light toward a detector
when no absorber is settled, while $I_{\rm abs}$ is the intensity when
one or multiple absorbers are embedded.
The temporal profiles of $A(t)$ provide training data in the LSTM machine learning. 
For an incident laser beam, we obtain datasets for 
temporal profiles of $A(t)$ at the positions of 8 detectors.

Here, we consider the case for a laser beam injected from S2. 
Fig. 2 shows the resultant temporal profiles of absorption measure $A(t)$,
depending on the position of absorber.
For example, the upper left panel shows temporal profiles of $A(t)$ for a absorber at \#3. 
Since the absorver is very close to detector D8, a curve labeled with D8 shows significant
absorption. 
Similarly, the lower left panel (absorber at \#0) and the upper right panel 
(absorber at \#26) exhibit strong absorption on the detectors 
in the vicinity of absorber.
Lower right panel (absorber at \#13)  shows the result for absorber 
near the center of the whole region.
In this case, all detectors recognize significant absorption, and also
temporal profiles depend upon the detector position. 
These features of absorption measure $A(t)$ are used for training data
in the present deep learning. 

\begin{figure}[h]
\centering
\includegraphics[width=8cm]{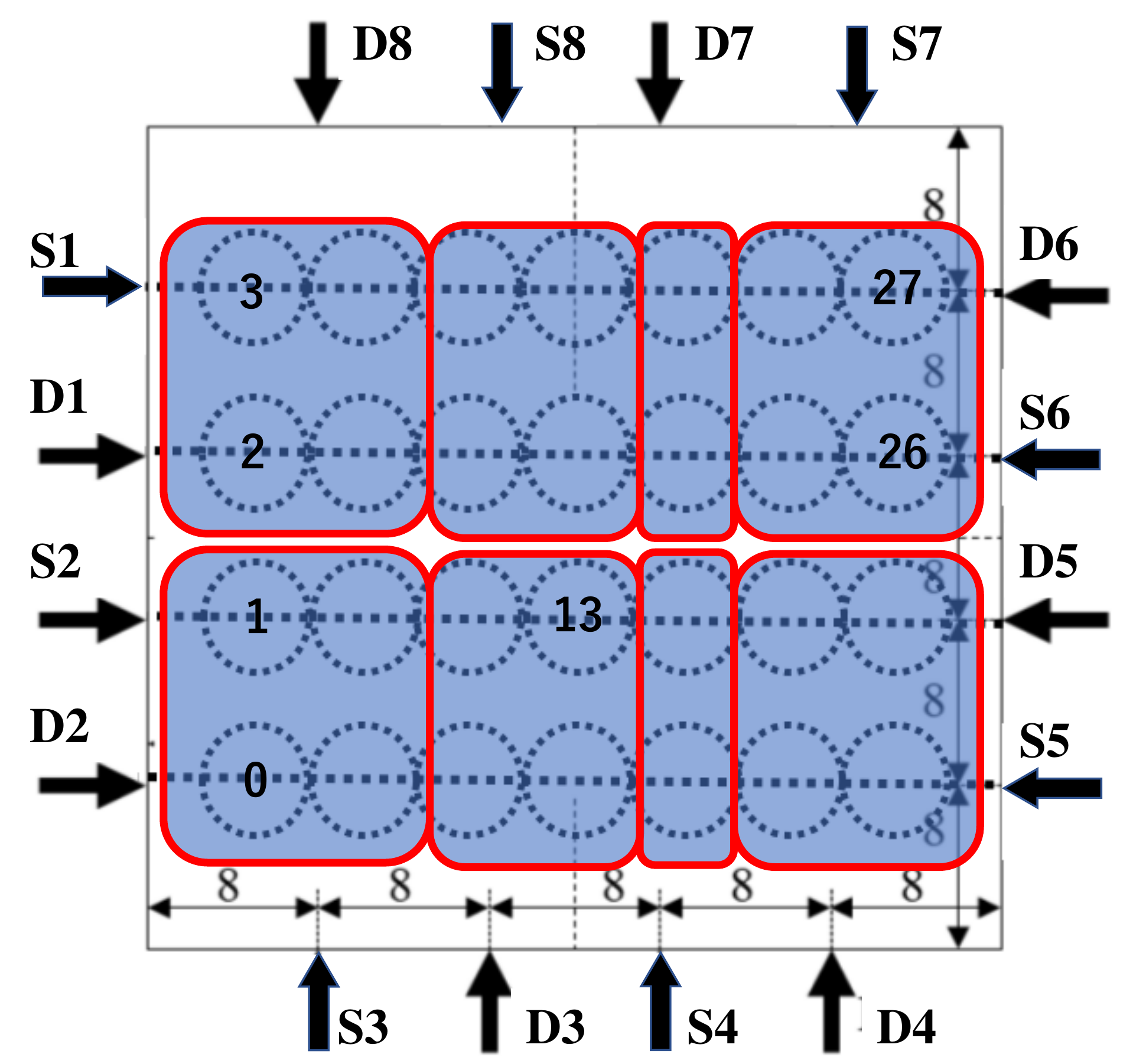}
\caption{The configuration of target tissue. The size is assumed to be 
4cm $\times$ 4cm. There are located 28 domains, each of which can 
possess a round absorber with diameter 5mm. 
8 incident beam directions (labelled S1-S8) and  
 8 detector positions (D1-D8) are shown. We divide the whole area into 8 groups,
 each of which has four or two domains. }
\end{figure}
\begin{figure}[h]

\centering
\includegraphics[width=12cm]{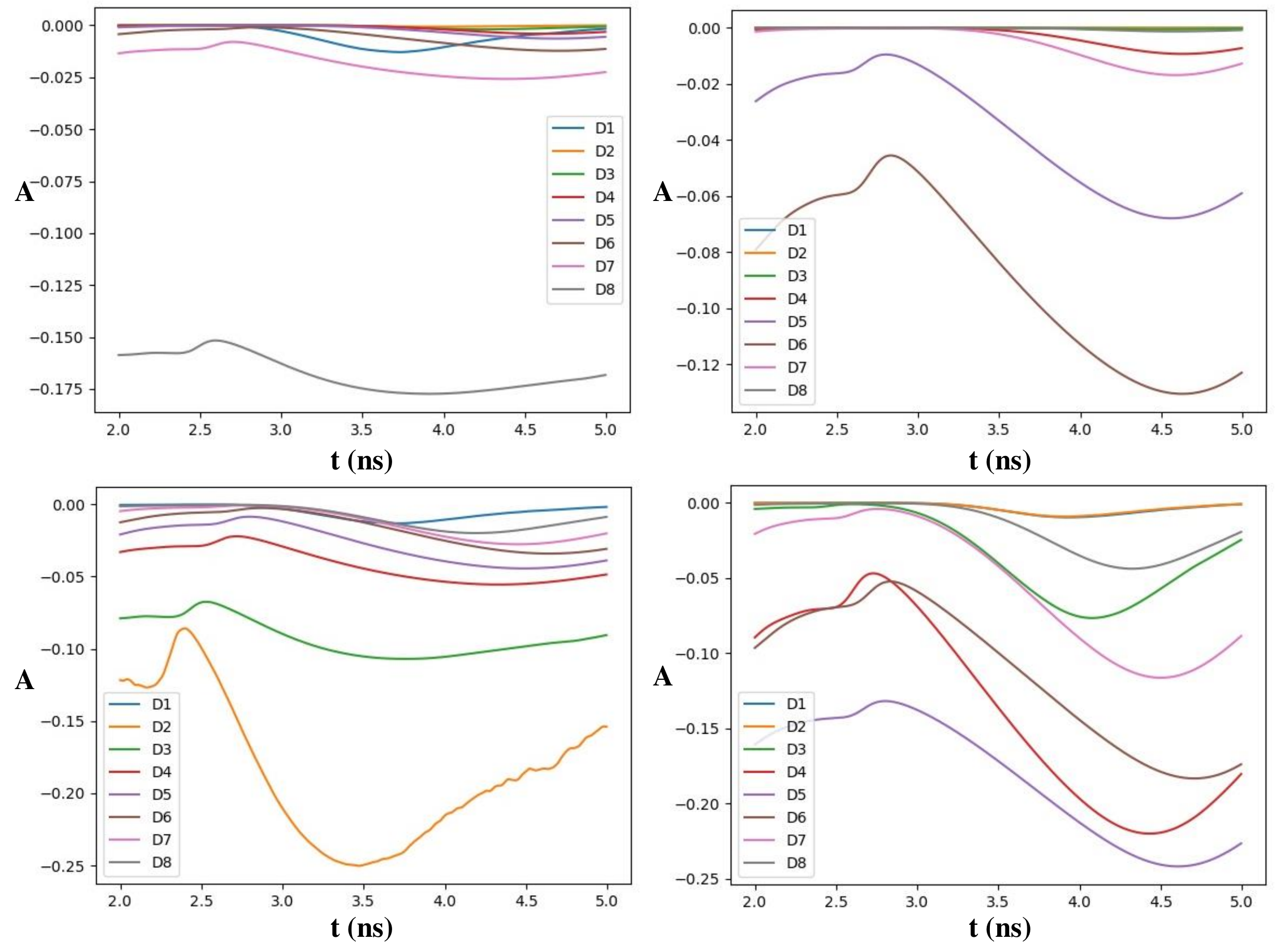}
\caption{Temporal profiles of absorption measure $A={I_{\rm abs}(t) /I_{\rm noabs}(t)}-1$ on 8 detectors (D1-D8).
The horizontal axis is time in units of nanoseconds (ns).
Upper left panel shows the profiles for an absorber located at  \#3 position,
upper right panel for an absorber at \#26 position,
lower left panel for an absorber at \#0 position, and
lower right panel for an absorber at \#13 position. }
\end{figure}

\subsection{LSTM deep learning method}

In order to classify absorber positions, we employ a deep learning method. 
Deep learning based on artificial neural networks is one of powerful 
machine learning  methods. 
Especially for temporal data, 
Long Short Term Memory (LSTM) learning is an effective tool 
that is an extension of artificial recurrent neural network (RNN) architecture 
to analyze not only single data points but also entire sequences of temporal data.
LSTM networks have improved recurrent neural network 
by using the gates to selectively retain and forget information, 
which are relevant and not relevant, respectively. 
Lower sensitivity to time gaps makes LSTM networks robust 
for analysis of temporal data than a simple recurrent network. 

We conduct deep learning based on two LSTM layers plus final Dense layer by using Tensor flow,  
which is an open platform for a machine learning 
provided by Google coorperation. 
In order to  classify multiple domains, we use a categorical cross entropy 
for a loss function and a softmax activation function. 
The cross entropy loss function is defined as 
\begin{equation}
-\sum_{c=1}^M y_{c} \log (p_{c}) \,,
\end{equation}
where $M$ is the number of domains,
$y_{c}$ is a binary indicator which takes $1$ if domain label $c$ is 
the correct classification otherwise $0$,
and $p_c$ is predicted probability observing domain $c$. 
We calculate a separate loss for each domain label and sum up over all domains. 
The softmax activation function is defined as 
\begin{equation}
f_c(x_c)=\frac{e^{x_c}}{\sum_{i=1}^M e^{x_i}}\,,
\end{equation}
where $x_i$ is an element of input data vector ${\bf x}$.
The softmax activation function is mainly used for a classification problem.

\section{Multi-step classification method}

It is difficult to classify all 28 positions directly at once.
Thus, we make two-step classification. 
In the first step, all dataset are classified into 8 groups of domains, and 
each group is classified into domains in the second step. 
We find that such multi-classification method is useful for absorber detection. 
In Fig. 1, two-step classification is illustrated. The small box represents 8 groups and one 
group is composed of 4 or 2 domains. 

The training and test data are the temporal profiles of absorption measure $A$
at 8 detector positions,
which are obtained by radiative transfer simulations for a absorber located
at a domain. 
Moreover, we add random noises on simulation data. 
We attempt two types of noises.
One is based on uniform random number in $(0,1)$ and 
its amplitude is assumed to be $0.01$ in absorption measure $A$. 
The other is Gaussian noises, where random number produced 
by a Gaussian distribution function, $\exp(-x^2/2\sigma^2)$ with a 
standard deviation $\sigma$. 
In the present analysis, we take $\sigma=0.001$. This noise level is 
one tenth of uniform random noises, but it is suitable for classification 
by {\sf LSTM} deep learning. 
Temporal profiles of absorption measure with noises are plotted in Fig. 3. 
We generate  2500 datasets in total,  which include
89 datasets for each absorber position. 

\begin{figure}[h]
\centering
\includegraphics[width=12cm]{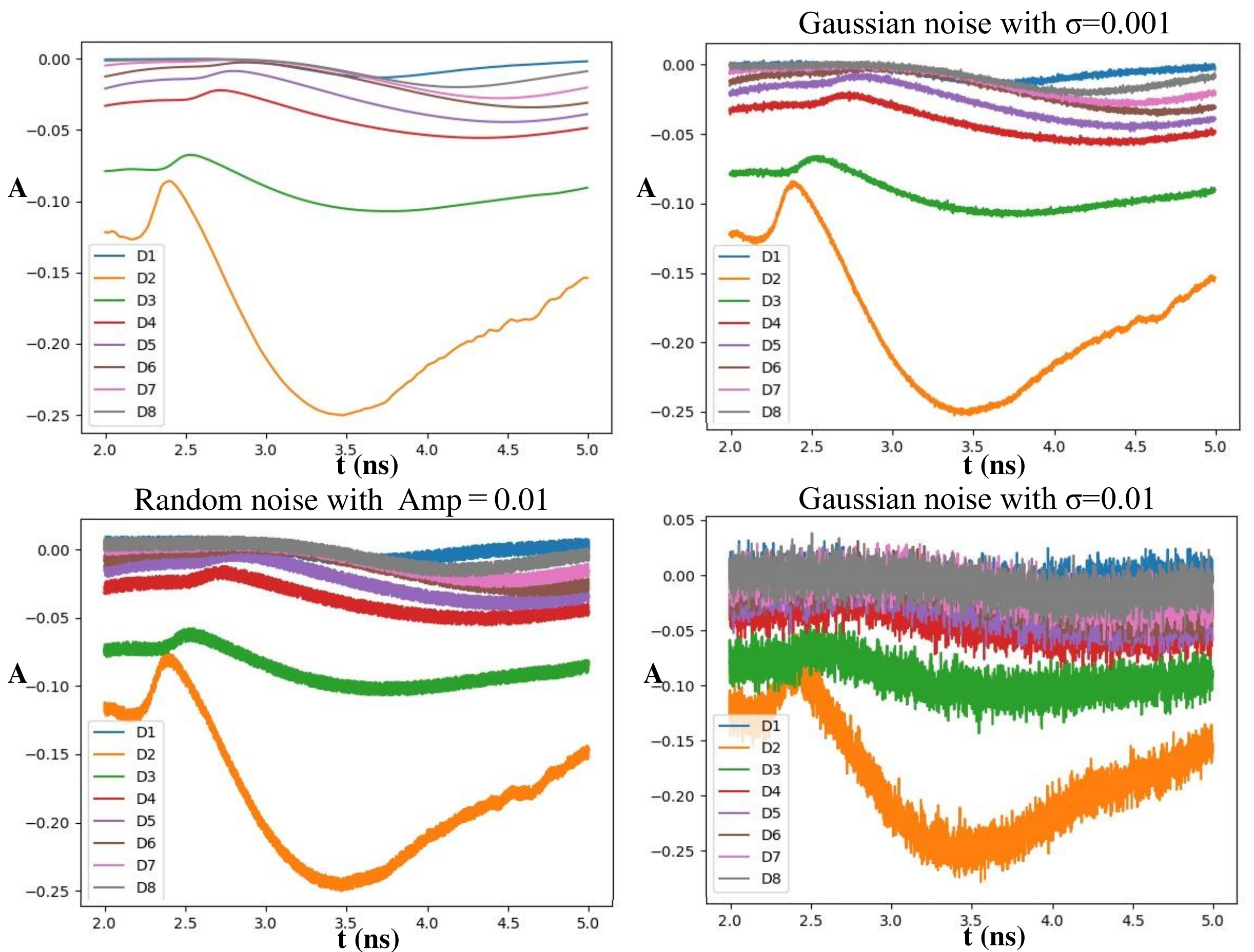}
\caption{Generation of datasets with noises: 
Upper left panel shows raw data by radiative  transfer simulations. 
Lower left panel shows training data for deep learning with uniform random noises whose amplitude is $0.01$. Upper right panel shows training data with Gaussian noises
whose standard deviation is $\sigma=0.001$, 
and lower right panel is test data with Gaussian noises of $\sigma=0.01$. }
\end{figure}

The first classification is constructed to determine learning parameters, 
where learning epochs are about 40 and the number of training data 
and test data is 2500 and 500, respectively. 
The final value loss is 0.03 and accuracy rate reaches 99\% at the final epoch. 
The sub-classification in the second step is achieved more easily.
For groups composed of  4 domains,
learning epochs are 10 and the number of training data and test data is respectively
1000 and 200.
For groups composed of  2 domains,
learning epochs are 10 and the number of training data and test data is respectively
600 and 100.
The final accuracy rate reaches 100\% for such second-step classifications. 

The constructed detection models can be tested using test data, which are 
generated by the same procedure for noise addition. 
First, we tried to test the detection model for uniform random noises. 
The resultant accuracy rate falls to $83\%$. It means that one always makes two  
 mistakes to get two wrong positions within the total positions  of 28. 
But, such discrepancy can be fixed for models with just one absorber. 
For the purpose, we reconstruct another dataset of training samples 
by using new random number. 
It predicts positions other than the first predicted position. 
When we use and compare predictions between two sets of detection models. 
the final accuracy rate reaches $96\%$ as a result of such double classification. 

If we employ Gaussian noises, the accuracy rate is much more improved.
The model with Gaussian noises can detect the correct position for test data 
with larger noise of $\sigma=0.01$, which is ten times larger than training data.  
Using test data with such Gaussian noises, the accuracy rate reaches $100\%$. 
If we use a model with uniform random noise of 0.01, then
the accuracy rate falls to $89\%$. 
It clearly demonstrates that noise types of test data are sensitively recognized
by {\sf LSTM} deep learning.
Hereafter we use the detection model based on test data with Gaussian noises. 

Also, we study the effects by offsets of absorbers and setups of finer domains.
Fig. 4 shows the results for an absorber located between domains.
An absorber is shown by a red circle and the predicted position is 
indicated by a square symbol. 
The upper left panel shows that
the present model can predict a position near the correct position. 
The upper right panel shows the results for an absorber at the middle of 
4 domains. Again, a near position is predicted.  
Next, we consider the refinement of domains. 
In the lower panels of Fig. 4, the results for 8 domains are shown. 
The prediction of a position near the correct position allow us to get 
more accurate information on the absorber position. 

\begin{figure}[h]
\centering
\includegraphics[width=8cm]{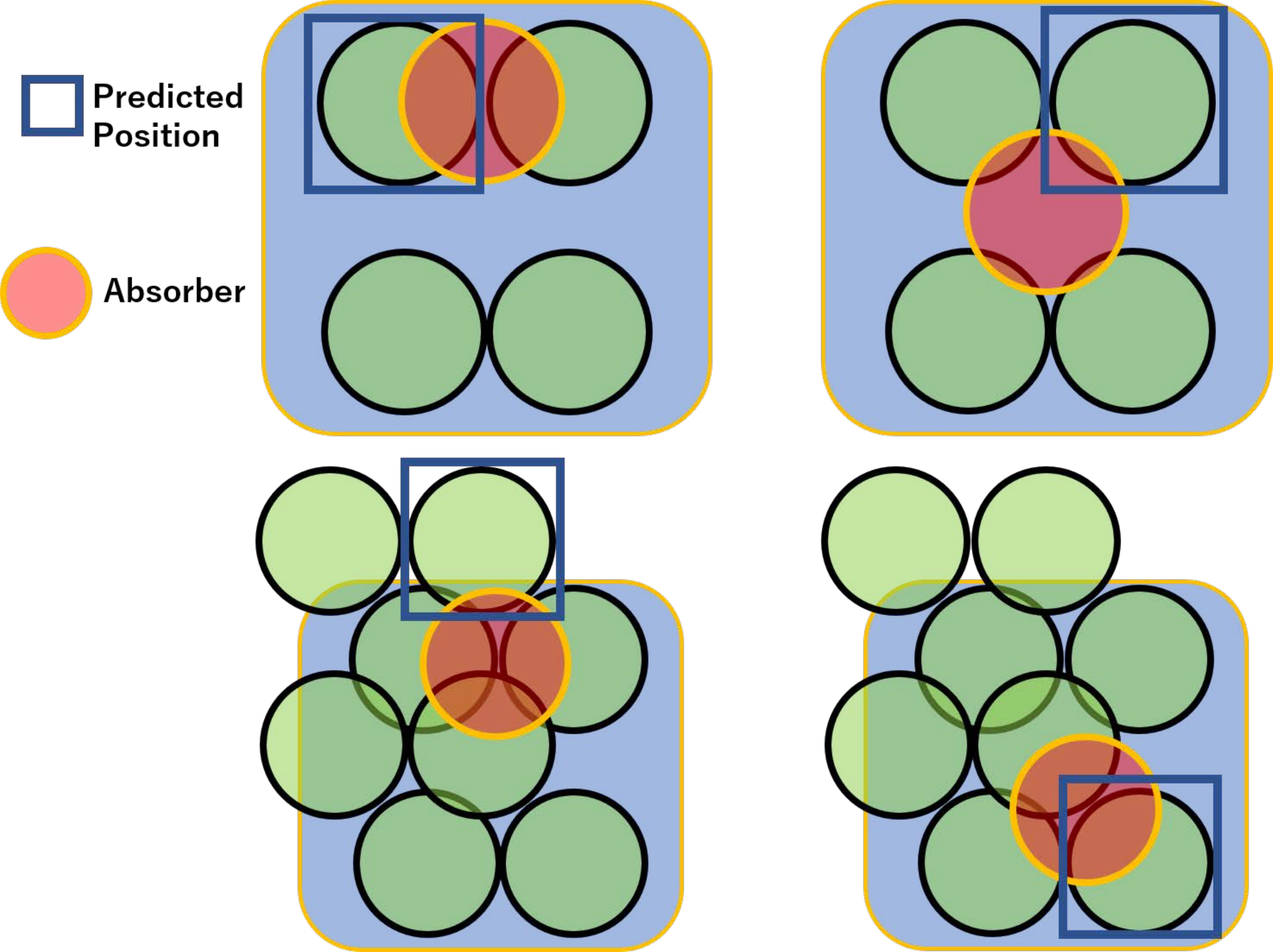}
\caption{Results of position prediction for absorbers with offsets.
Red circles are absorbers, and blue squares are predicted positions.
}
\end{figure}


\begin{figure*}[t]
\centering
\includegraphics[width=12cm]{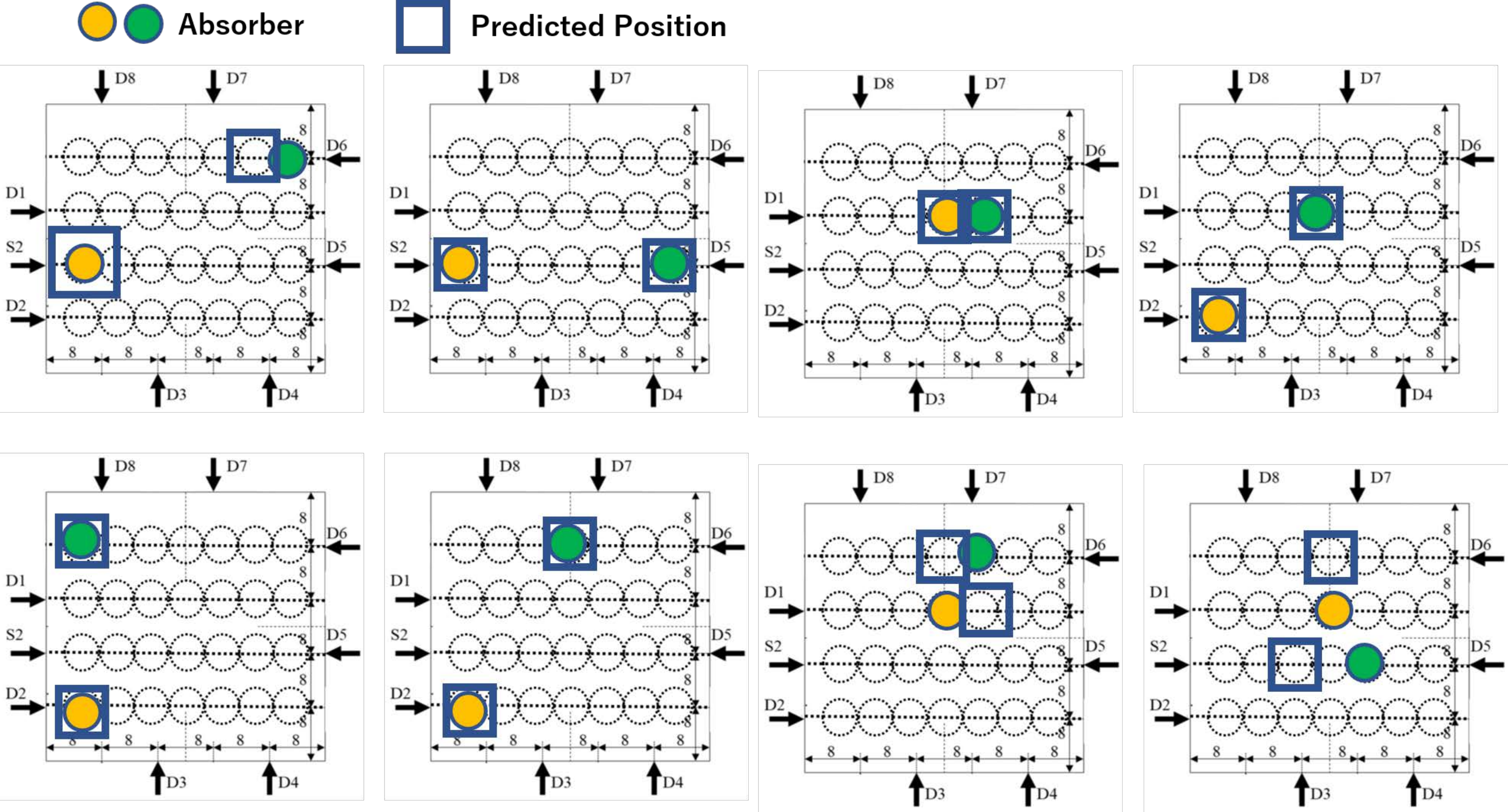}
\caption{The results for two absorbers. Orange and green circles
are absorbers, and predicted positions are indicated by blue squares.}
\end{figure*}

\begin{figure}[h]
\centering
\includegraphics[width=10cm]{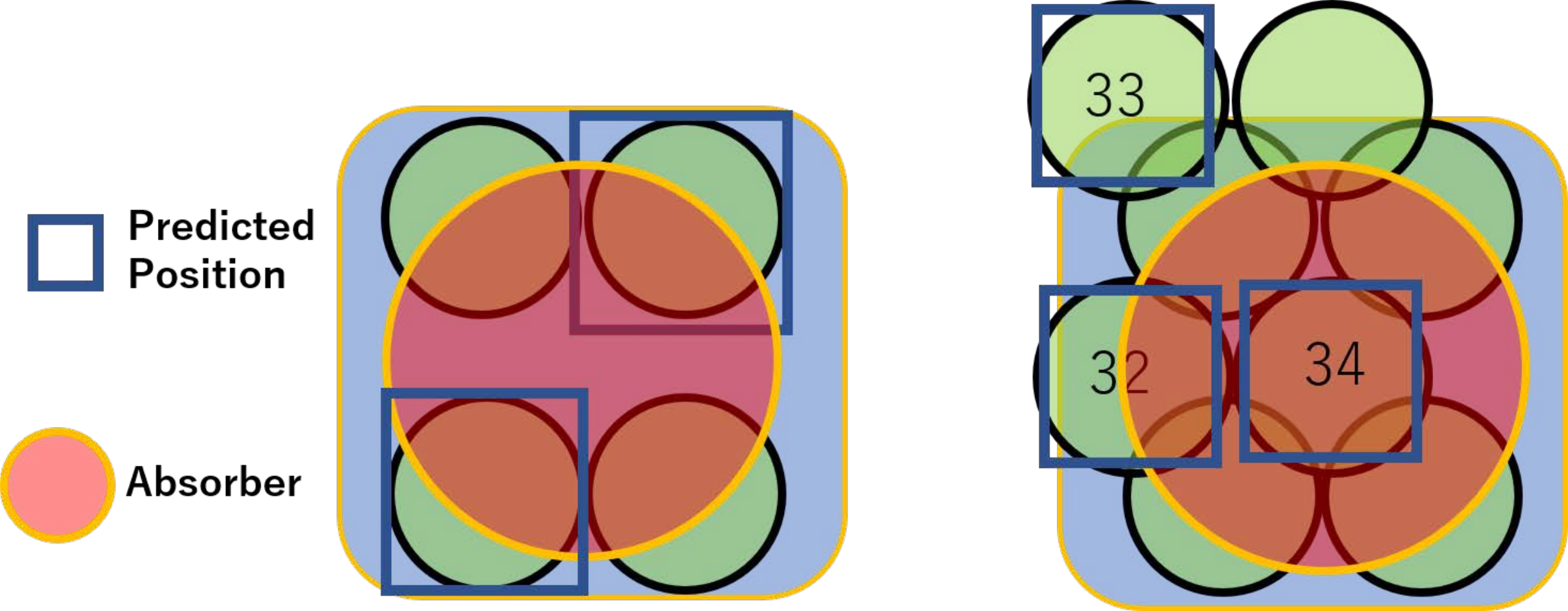}
\caption{The results for a big absorber. 
The left is the classification in 4 domains, while
the right is that in 8 domains.
A red circle is an absorber, and predicted positions are indicated by blue squares.}
\end{figure}

\section{Data subtraction method}

Here, we demonstrate the effectiveness of data subtraction
to detect two absorbers or an absorber bigger than a domain.  

\subsection{Two absorbers}

First, we obtain temporal profiles of absorption measure 
by radiative transfer calculations for absorbers located at two domains. 
Then, we apply  our model with training data,  
and make prediction with test data. 
Once a position of absorber is predicted, 
we subtract the profile data for the absorber from the original data. 
Then, we apply again our detection model to the data after subtraction. 
In this subtraction method, we can successfully predict positions of two absorbers. 
Fig. 5 shows the results for two absorbers, where orange and green circles
are absorbers and predicted positions are indicated by blue squares.
In most cases, correct positions or near positions are predicted, 
while in a lower right case, prediction is not so precise.
After we tested 60 cases for two absorbers, 
we found that the data subtraction method works well to
give a high accuracy rate. 

These results imply that a linear combination of each absorption profile 
is a good approximation for most cases of two absorbers.
When the distance between two absorbers is long,  the linearity makes sense.
But, the upper middle panel in the Fig. 5 shows 
that the linearity is a good approximation even for adjacent absorbers. 
This is  the case where two absorbers are located in parallel to  the incident beam direction.
However, in the lower right panel, a nonlinear effect by two absorbers 
seems to be important. 

\subsection{Bigger absorbers}

Here, we consider an absorber two times bigger than a domain. 
Temporal profiles of absorption measure for this absorber is obtained
by radiative transfer calculations. 
Then, our detection method is adopted for the data.
Fig. 6 shows the results of prediction for a big absorber.
The left panel of Fig. 6 shows the classification 
based on the subtraction method in  4 domains. 
Two domains adjacent  to the absorbers are predicted  with this method.
The right panel of Fig. 6 shows the classification in 8 domains.  
In this case, the machine learning without data subtraction gives us the probability 
distributions of correct detection.
The resultant probabilities for the position: 
$72\%$ for \#34, $16\%$ for \#33,  and  $10\%$ for \#32. 
Such probability distributions are informative for a big absorber. 
We also suppose the case of 1.5 times bigger absorber, and
the resultant probabilities for the position: 
$75\%$ for \#34, $13\%$ for \#33, and $8\%$ for \#32. 
Compared to 2 times bigger absorber, probabilities at  \#32 and \#33
are slightly smaller. 
Thus, these probability distributions reflect the size of absorber. 

As shown above, the data subtraction method can predict the correct position
or nearest ones. 
In order to evaluate the probability of correct detection,
it is useful to introduce the scoring rule.
As shown in the left panel of Fig. 7, the score of correct position is 2,
while the score of adjacent positions is 1.
The right panel of Fig. 7 shows the example of total scores.
If the position of one absorber is correctly predicted, the score is 2,
while the score is 1 if an adjacent position of another absorber is predicted.
Thus, the total score is 3, which means the detection probability of $75\%$.
Based on the scoring rule, we tested 60 examples for two absorbers 
and got the detection probability of $72\%$ on average. 
It means that the most probable prediction for two absorbers is
the combination shown in the right panel of Fig. 7, where
one absorber is correctly predicted and an adjacent position of 
another absorber is predicted. 

\begin{figure}[t]
\centering
\includegraphics[width=8cm]{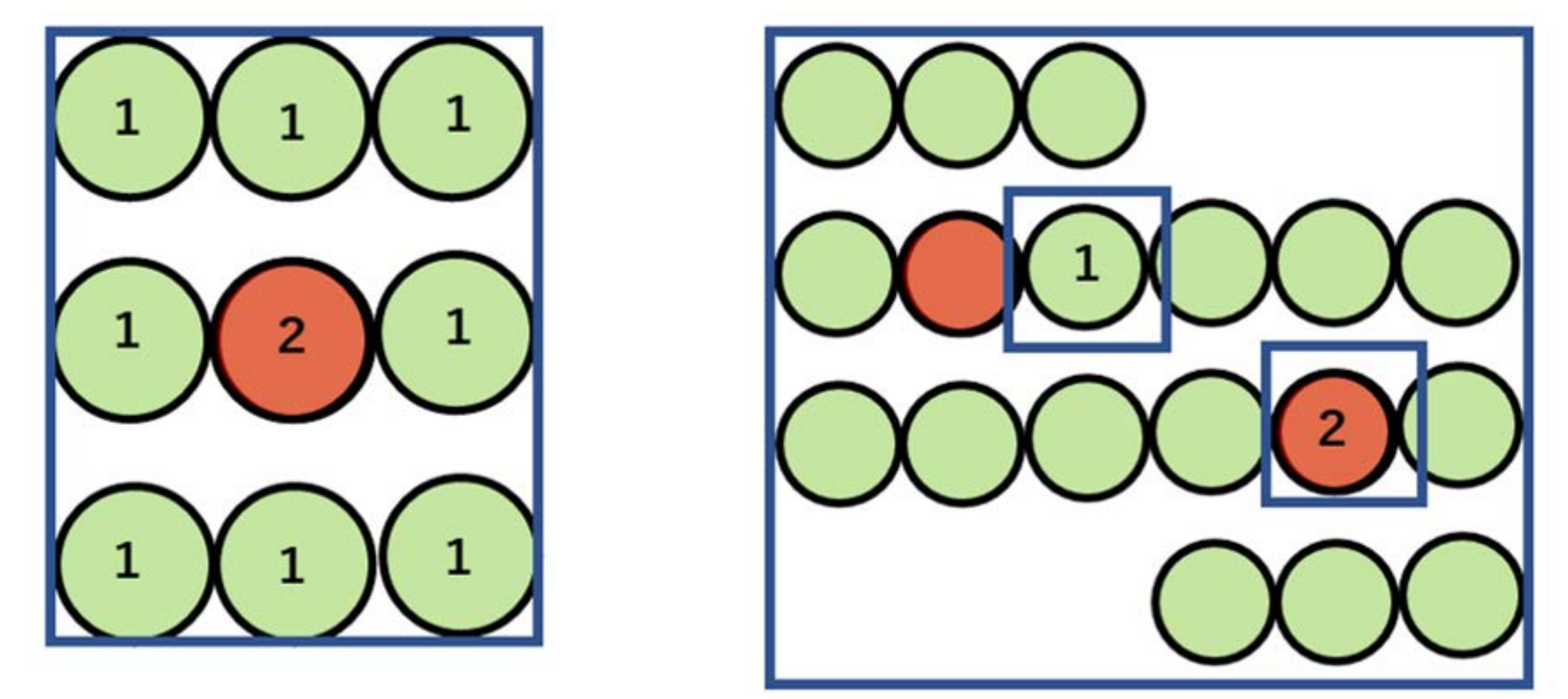}
\caption{The left panel shows a point at the position for our scoring. If one detect the absorber position, one  get 2 points. The right panel shows one example of scoring for two absorbers, that leads to total 3 points, which means the detection probability of $75\%$.}
\end{figure}

\begin{figure}[ht]
\centering
\includegraphics[width=12cm]{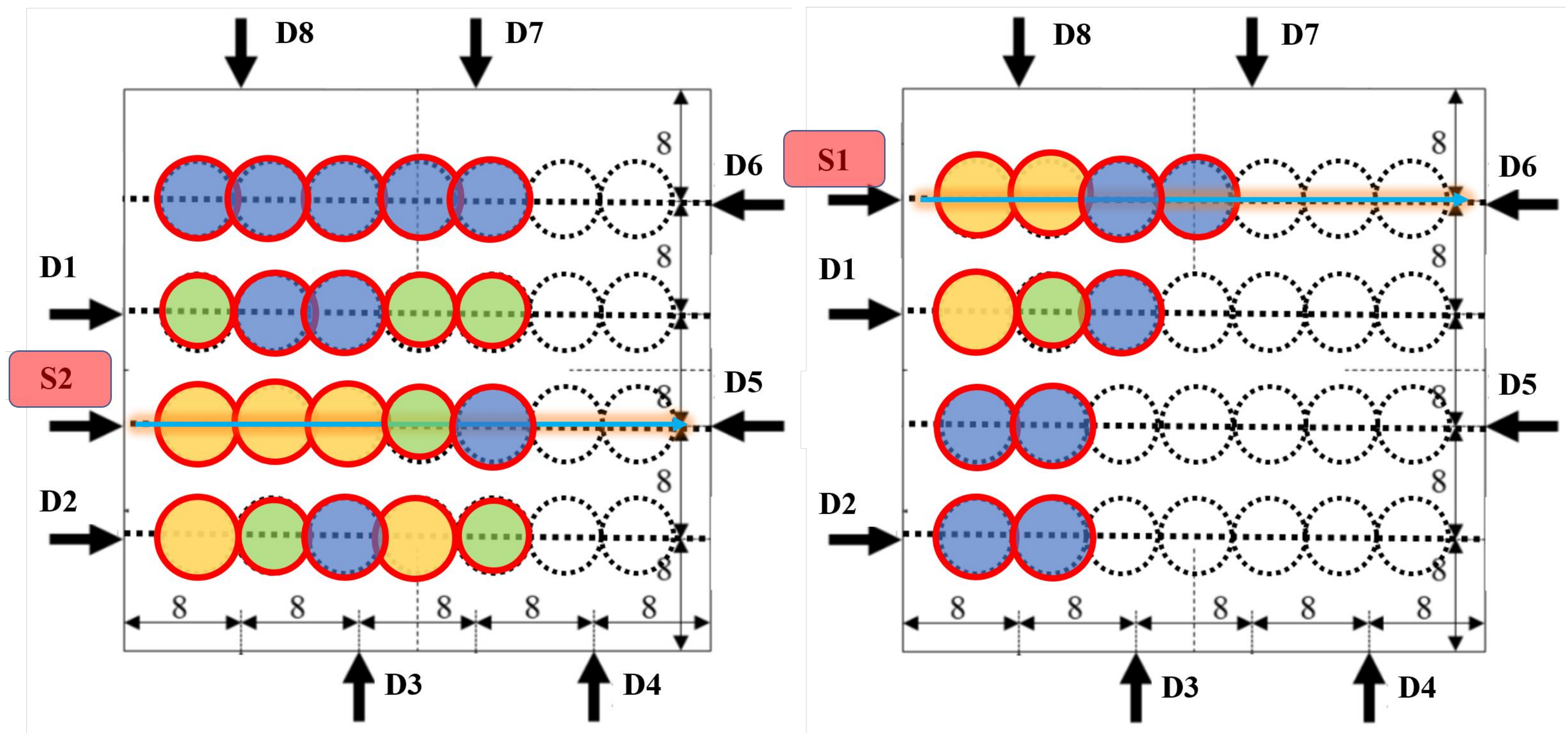}
\caption{Distributions of detection probabilities 
for an incident beam from S2 (left panel) and S1 (right panel): 
yellow domains are most probable positions for detection, 
while green domains correspond to medium probabilities
and blue domains do to low probabilities. }
\end{figure}
\begin{figure}[h]
\centering
\includegraphics[width=8cm]{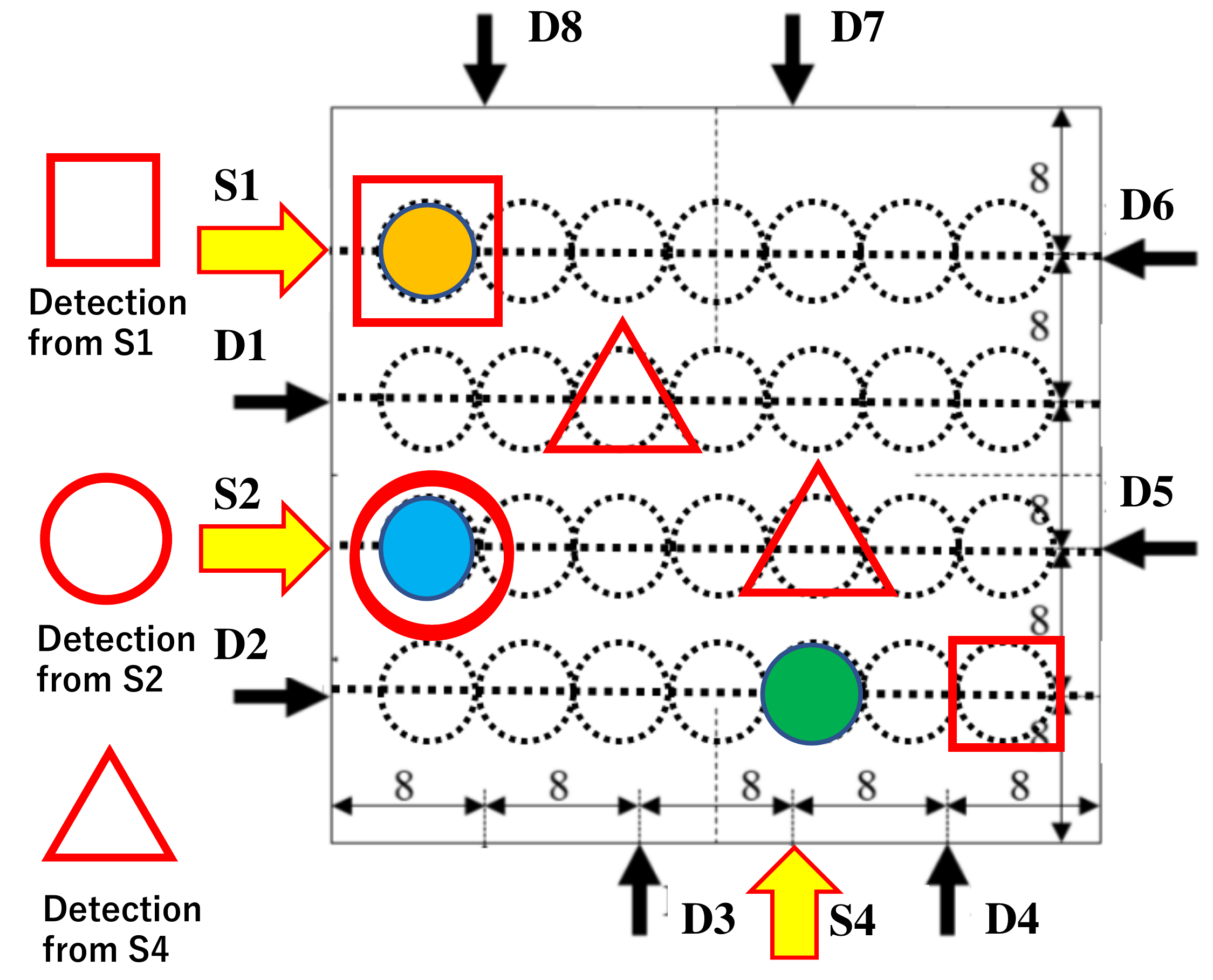}
\caption{Example for multi-beam injection model.
Three absorbers are located, and three different beams are injected. 
Each symbol corresponds to detected positions for each beam.
}
\end{figure}

\section{Multi-beam injection model}

So far we considered one incident laser beam from S2 in Fig. 1. 
Here, we analyze the case where laser beams are injected from multiple positions, 
that is, S1 to S8 in Fig. 1. We can use 8 datasets for different incident beams. 
In the data subtraction method, it is found that 
the detection probability is higher for an absorber located near an incident beam.

Fig. 8 shows the distributions of detection probabilities. In yellow domains that are
near the beam position, detection probabilities are high,
while probabilities are low in blue domains.  
If we apply the present method to detasets with Gaussian noises
for three incident beams from S1, S2 and S4, 
the accuracy rate results in 100$\%$ to detect one absorber.
Also, we tested the multi-beam model for three absorbers, using data subtraction method.
The results are shown in Fig. 9, where
each symbol corresponds to detected positions for each beam.
In this example, the blue absorber is correctly detected, while 
the orange absorber is within two detected positions indicated by squares. 
The total score is 7, where 4 for S2, 2 for S1 and 1 for S4, to the full score of 12. 
Hence, this result means accuracy rate of $7/12=58\%$. 
If additional datasets are given for different beams, it is expected to improve 
the accuracy rate for detection.

\section{Discussion}

\subsection{Validity of 2D RTE}

In this paper, we have used 2D simulation data 
based on a radiation transfer solver {\sf TRINITY}. 
Here, we perform  3D Monte Carlo simulations to check  the validity of 2D models. 
We pursue the propagation of $10^{12}$ photon packets, each of which 
is injected in random directions from a laser beam point. 
Each photo packet travels up to its optical depth defined by random number $R$
as $\tau=- \log(R)$. Fig. 10 shows the absorption measure $A$ 
calculated by Monte Carlo simulations. 
Although the profile is quite noisy, 
it can be integrated over some time interval, that 
is, a green line in the figure. This feature can be traced by 2D radiative transfer simulations. 
If we tentatively apply the Monte Carlo simulation data to
our machine learning model,  we can predict an exact position of absorber. 

\subsection{Sampling effect}

In the present analyses, we utilized datasets of all time steps for machine learning.
Actually, a time step in our simulations is $\Delta t=0.0006$[ns], while 
it is $\Delta t=0.01$[ns] in  experiments using the phantom.
One advantage of LSTM deep learning is to recognize global behaviors of temporal data.
To check this advantage, we tentatively conducted the same LSTM deep learning
using $1/16$ data points (16 times longer time steps).
As a result, we have confirmed that the present LSTM model does work
even for such coarse-grained data. 
So, this advantage allows us to speed up machine learning dramatically,
and to adjust the time step to that of phantom experiments.

\begin{figure}[h]
\centering
\includegraphics[width=8cm]{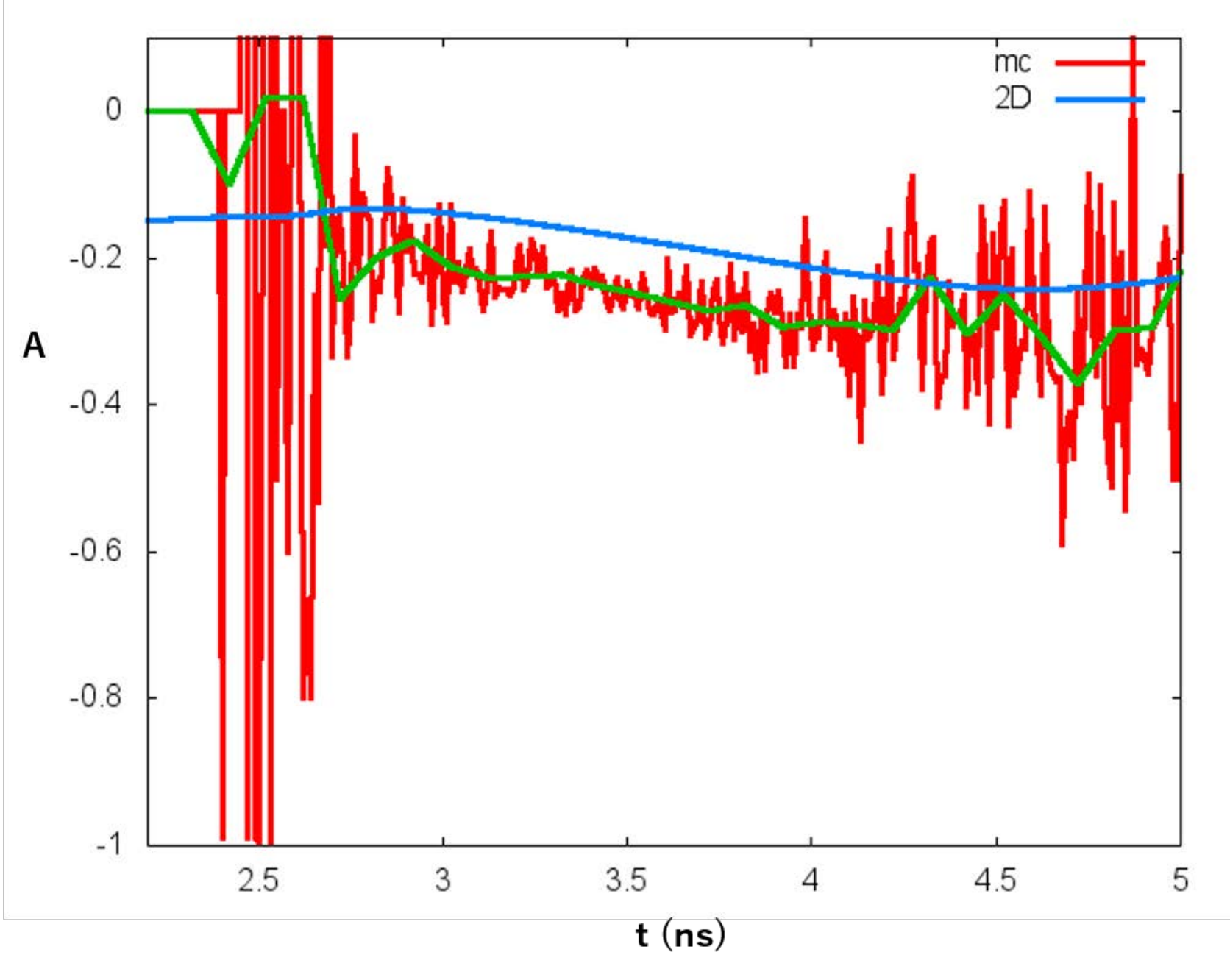}
\caption{Absorption measure at the detector D5 calculated by Monte Carlo simulations.
A green curve is a smoothed profile of raw data.  
A blue curve is the result by 2D radiation transfer simulations. 
 }
\end{figure}

\section{Summary}  

In this paper, we have explored a novel deep learning algorithm based on time-domain radiative transfer equation.
To specify positions of cancer cells (absorbers), we have applied a long-short-term memory (LSTM) deep learning method to temporal profiles of absorption measure for a two-dimensional model of target tissue, which are calculated by solving time-domain radiative transfer equation.

We have shown that positions of absorbers
can be predicted with high accuracy rates
by a multi-step classification method. 
We attempted two types of noises on simulation data 
and investigated resultant accuracy rates.
As a result, we found that LSTM deep learning can recognize sensitively noise types of test data. 

We have also developed methods to detect two absorbers or an extended absorber larger than a domain. We demonstrated that data subtraction method successfully works to predict nearly correct positions of absorbers.

The present method is based on the assumption of linear combination of each absorption profile. However, some examples show that nonlinear effects are important.
In a future work, we will attempt to extract nonliner effects 
to raise the accuracy rate.
Also, we tested a multi-beam model for three absorbers and showed the total accuracy rate is $58\%$. If additional datasets are given for different beams, we expect the accuracy rate of detection can be improved. We will construct a more sophisticated model based on many beams in a future work.

\section*{Acknowledgments}
This research is supported by Department of Computational Medical Science in Center for Computational Sciences, University of Tsukuba.

\end{document}